\documentclass[aps,pre,twocolumn,showpacs,notitlepage,floatfix,superscriptaddress]{revtex4-1}

\usepackage{amsmath}
\usepackage{amssymb}
\usepackage{mathtools}
\usepackage[all]{xy}
\input xy
\xyoption{matrix}
\xyoption{2cell}
\xyoption{arrow}
\xyoption{curve}
\xyoption{all}

\usepackage{graphicx}
\usepackage{epsfig}
\usepackage{dcolumn}
\usepackage{bm}

\usepackage{amsmath}
\usepackage{graphicx}
\usepackage[latin1]{inputenc}
\usepackage{ulem}
\usepackage{epstopdf}
\usepackage{subfigure}
\usepackage{color}

\begin{document}
\title{ Quasiperiodic driving of  Anderson localized waves in one dimension}

\author{H. Hatami}
\affiliation{Center for Theoretical Physics of Complex Systems, Institute for Basic Science, Daejeon, Korea}
\author{C. Danieli}
\affiliation{New Zealand Institute for Advanced Study, Centre for Theoretical Chemistry \& Physics, Massey University, Auckland, New Zealand}
\author{ J. D. Bodyfelt}
\affiliation{New Zealand Institute for Advanced Study, Centre for Theoretical Chemistry \& Physics, Massey University, Auckland, New Zealand}
\author{S. Flach}
\affiliation{Center for Theoretical Physics of Complex Systems, Institute for Basic Science, Daejeon, Korea}
\affiliation{New Zealand Institute for Advanced Study, Centre for Theoretical Chemistry \& Physics, Massey University, Auckland, New Zealand}

\date{\today}

\begin{abstract}

We consider a quantum particle in a one-dimensional disordered lattice with Anderson localization, in the presence of multi-frequency perturbations of the onsite energies. 
Using the Floquet representation, we transform the eigenvalue problem into a Wannier-Stark basis.
Each frequency component contributes either to a single channel or a multi-channel connectivity along the lattice, depending on the control parameters. 
The single channel regime is essentially equivalent to the undriven case. 
The multi-channel driving substantially increases the localization length for slow driving, showing two different scaling regimes of weak and strong driving, yet the localization length stays finite
for a finite number of frequency components. 


\end{abstract}
\maketitle
\section{Introduction}

Disordered systems have been of large interest for transport phenomena since Anderson predicted that waves localize in the presence of uncorrelated random potentials \cite{Anderson58}. By obtaining the exponential decay of every eigenstate of a disordered class of one dimensional linear wave equations, Anderson proved the localization of a single quantum particle, within a finite volume of the chain, due to disorder. Ever since then, theoretical studies have been performed on higher dimensional lattices in random potentials, showing absence of diffusion in the two dimensional case \cite{Abrahams79}, and mobility edges between the insulating and the metallic phase in the three dimensional case \cite{Bulka85}. In one dimension, Anderson localization has been observed in experiments with light waves \cite{Lahini08} and atomic Bose-Einstein condensates \cite{ Billy08, Roati08}, enhancing further interest in theoretical research of disordered systems.

Wave localization inherently relies on the phase coherence within the wave state. If the disordered potential is allowed to temporarily fluctuate in a random way, phase coherence is lost, and 
the previously localized wave starts to diffuse without limits \cite{Rayanov13}. Assuming that the temporal fluctuations are represented as a quasiperiodic function of time with $D$ incommensurate fundamental frequencies, the random noise can be effectively reached in the limit of $D \rightarrow \infty$. Here we address the case of a finite number of frequencies (colors) $D$.
Will the localization length $\zeta$ stay finite for any finite $D$? If yes, what is its dependence on $D$? At fixed $D$ how does $\zeta$ depend on the remaining control parameters, such as frequency, fluctuation amplitude and disorder strength? 

A series of computational studies was devoted to this very issue \cite{Yamada93,Yamada98,Yamada99}. While the first conclusion was that Anderson localization can be destroyed for
$D\geq 2$, a more accurate recomputation showed that the localization length might well increase, yet stay finite. Other studies focused on the case of single color $D=1$  and computed conductance
through small finite systems, without clear conclusions on the localization length \cite{Martinez06, Kitagawa12}.

In this work we first consider a single frequency color, that is a time-periodic drive. We use the Floquet representation to arrive at a time-independent eigenvalue problem on a two-dimensional
lattice, with one direction corresponding to the original spatial extension, and the second one to the Floquet (driving) one. We transform into a Wannier-Stark basis which is diagonal along
the Floquet direction, and analyze the resulting eigenvalue problem. For large driving frequencies the equations reduce to uncoupled single channel ones, which are essentially equivalent to
the undriven case. For small driving frequencies we obtain a multi-channel regime with a substantial increase of the localization length, and its divergence in the limit of vanishing frequency.
This multi-channel regime divides into two further regimes of weak and strong driving amplitudes, which yield different scaling laws.
We then generalize to the case of many incommensurate frequencies, and compare our findings to numerical results.

The paper is organized as following. In Sec. \ref{sect:Intro} we introduce the model and its general features. In Sec. \ref{sect:D=1} 
we derive the results for one frequency (color) drive. We generalize to many colors in Sec. \ref{sec:General case}, and discuss numerical results in Sec. \ref{sect:wave dyn}. We conclude with
discussions, a summary, and an outlook. 

\section{Model} \label{sect:Intro}
We consider a disordered one-dimensional tight-binding chain in the presence of a coherent time-dependent driving of the onsite energies. The equations of motion read
\begin{equation}
i\dot{\psi_l}=\epsilon_l\bigg[1+\sum_{i=1}^D\mu_i \cos(\omega_i t+\phi_l^i)\bigg]\psi_l-\lambda(\psi_{l+1}+\psi_{l-1})\ ,
\label{Eq:multicolor}
\end{equation}
The onsite energies of each lattice site $\epsilon_l$ are random uncorrelated numbers with a probability density function (PDF) of value $1/W$ inside the interval $\epsilon_l \in [-W/2,+W/2]$ and 
zero outside. The parameter $W$ parametrizes the strength of the disorder. The coefficient $\lambda$ is the strength of the hopping between nearest neighbor lattice sites, while $\mu_i$ and $\omega_i$  respectively are the amplitude and frequency of the $i$-th driving. $D$ is the total number of frequencies (colors) in the driving. The frequencies $\Omega = (\omega_1,\dots,\omega_D)$ are chosen incommensurate with each other
\begin{equation}
{\mathbf k\cdot \Omega}= k_1\omega_1 + \dots + k_D\omega_D\neq 0\ ,\quad \forall\ {\bf k}\in\mathbb{Z}^D\setminus\{0 \}\ .
\label{eq:incommens}
\end{equation}
The random phases $\phi_l$ are uncorrelated and have a PDF of value $1/(2\pi)$ inside the irreducible interval $\phi_l \in [-\pi,\pi]$. Their presence ensures broken time reversal symmetry of
(\ref{Eq:multicolor}).
For $\mu_i=0$, Eq.~(\ref{Eq:multicolor}) reduces to the well-known Anderson model with all eigenstates being localized with a finite upper bound on the localization length 
\cite{Anderson58,Kramer93}.  

\section{One color} \label{sect:D=1}
We first consider a driving with only one frequency $D=1$. In this case, Eq.(\ref{Eq:multicolor}) reads
\begin{eqnarray}
i\dot{\psi_l}=\epsilon_l\left(1+\mu \cos(\omega t+\phi_l)\right)\psi_l-\lambda(\psi_{l+1}+\psi_{l-1})~.
\label{Eq:onecolor}
\end{eqnarray}
Since the perturbation is time periodic with period $T=\frac{2\pi}{\omega}$, we first perform a Floquet expansion which will yield an effective two-dimensional lattice problem.

\subsection{From Floquet to Wannier-Stark} 
According to the Floquet theorem \cite{Floquet83, Shirley65}, a solution of (\ref{Eq:onecolor}) is given by
\begin{equation}
\psi_l (t)= u_l(t) e^{-iEt}~,
\label{eq:tr1}
\end{equation}
where $E$ is the quasienergy and the Floquet functions $u_l(t)=u_l(t+T)$. They can be represented in a Fourier series
\begin{equation}
u_l(t) = \sum_k A_{l,k} e^{i k (\omega t + \phi_l)} \;.
\label{eq:fourier transf}
\end{equation}
The Floquet expansion of the wave function $\psi_l(t)$ is then given by
\begin{equation}
\psi_l (t)=  \sum_k A_{l,k} e^{-i[(E- k\omega ) t - k \phi_l]} \;.
\label{eq:tr2}
\end{equation}
This transformation maps Eq.(\ref{Eq:onecolor}) into a time independent eigenvalue problem on a two-dimensional lattice (see appendix)
\begin{equation}
\begin{split}
E A_{l,k}& =( \epsilon_l + k \omega) A_{l,k} + \frac{\mu\epsilon_l}{2}\big(  A_{l,k-1} +  A_{l,k+1}  \big)\\
&-\lambda (\xi_{l,k}^{-}  A_{l-1,k}   +  \xi_{l,k}^{+}  A_{l+1,k})
\end{split}
\label{eq:AMrot1}
\end{equation}
with the coefficients
\begin{equation}
\xi_{l,k}^{\pm} = e^{-i k(\phi_l -\phi_{l\pm 1})} = e^{-ik\theta_l^{\pm}}
\label{eq:complex phase coefficients}
\end{equation}
dependent on the random phase differences $\theta_l^{\pm} = \phi_l -\phi_{l\pm 1}$, which introduce a synthetic gauge field in the two dimensional lattice of Eq.(\ref{eq:AMrot1})

Consider first $\lambda=0$. We can solve the remaining eigenvalue problem for each lattice site $l$ independently, as this case corresponds to the well-known 
Wannier-Stark ladder \cite{Fukuyama73} under an effective DC electric field $\omega$ and $l$-dependent hopping coefficient $\mu\epsilon_l/2$.
The eigenfunctions $B^{(\nu)}_{l,k}=J_{k-\nu}(\mu\epsilon_l/\omega)$ are obtained using the Bessel function of the first kind $J_{k}(x)$ with fixed argument $x$, and their eigenvalues form equidistant spectra $E_{\nu,l}=\epsilon_l + \omega \nu$ \cite{Watson22, Abramowitz72}. These eigenfunctions are localized (along the Fourier direction $k$), with tails which decay superexponentially 
fast. The localization volume (size) $\mathcal{L}$ of an eigenstate is estimated as $\mathcal{L} \sim 2\left| \frac{\mu \epsilon_l}{\omega} \right|$ for
$\frac{\omega}{\mu \epsilon_l} \leq 10$, and reaches its asymptotic value $\mathcal{L}=1$ for $\frac{\omega}{\mu \epsilon_l} \geq 10$ \cite{Krimer09}.

We use these Wannier-Stark eigenstates as a new basis for each lattice site $l$ of (\ref{eq:AMrot1}) :
\begin{equation}
A_{l,k} = \sum_{\nu} c_{l,\nu}  B_{l,k}^{(\nu)}\ .
\label{eq:tr3}
\end{equation}
The transformed eigenvalue problem for $\lambda \neq 0$ reads as (see appendix)
\begin{equation}
\begin{split}
&E c_{l,\nu} = (\epsilon_l + \omega\nu)c_{l,\nu} \\
-\lambda   \sum_{s} 
\Big[ & {\rm e}^{is\varphi_l^-}  J_{ s}\big(  \Delta_l^{-}  \big)  c_{l- 1,\nu-s} +  {\rm e}^{is\varphi_l^+}   J_{ s}\big(  \Delta_l^{+}  \big)  c_{l+ 1,\nu-s} \Big]
\end{split}
\label{eq:AM_newform}
\end{equation}
where
\begin{equation}
\begin{split}
&\tan \varphi_l^{\pm}  = -\frac{\epsilon_{l\pm1} \sin \theta_l^{\pm}}{\epsilon_l - \epsilon_{l\pm1}\cos \theta_l^{\pm}} \;, \\
&\Delta_l^\pm  = \frac{\mu}{\omega}\sqrt{ \epsilon_l^2 + \epsilon_{l\pm1}^2 - 2\epsilon_l \epsilon_{l\pm1}\cos \theta_l^\pm   } \;.
\end{split}
\label{eq:sqrt_Rphases}
\end{equation}
\begin{figure}[h]
      \includegraphics[width=0.95\columnwidth]{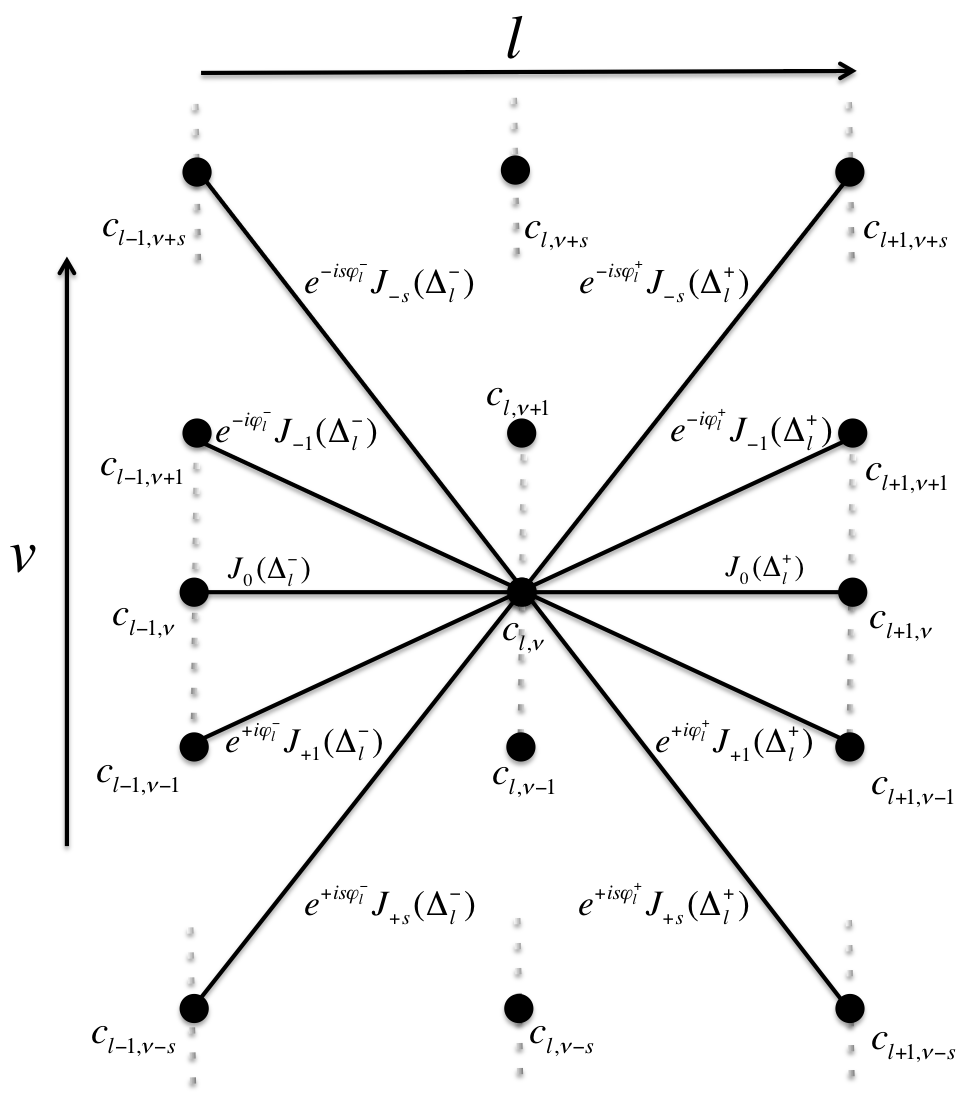}
    \caption{Two-dimensional eigenvalue problem Eq. (\ref{eq:AM_newform}). 
    Each given site $(l,\nu)$ is connected to a set of sites $(l\pm1,\nu -s)$ through the hopping $e^{is\phi_l^\pm}J_s(\Delta_l^\pm)$.} 
    \label{fig:Stark_Ladder2}
\end{figure}

This eigenvalue problem has zero hoppings along the Fourier direction $\nu$. Instead, each lattice site $c_{l,\nu}$ is connected to a number of nearest neighbor sites $c_{l\pm1,\nu-s}$ given
by the complex hopping coefficients $ {\rm e}^{is\phi_l^{\pm}} J_s(\Delta_l^\pm )$ (see Fig.~\ref{fig:Stark_Ladder2}).  The number 
of strong links  (connectivity) with a given site at $l,\nu$ depends on 
the ratio of the difference in the new onsite energies $| \epsilon_l - \epsilon_{l\pm1} + s \omega|$ and the hopping strength $|J_s(\Delta_l^\pm) | $. Strong links are then 
characterized by $| (\epsilon_l - \epsilon_{l\pm1} + s \omega) / J_s(\Delta_l^\pm) |\ \leq \ 1 $.

\subsection{Single channel vs. multi-channel regimes} 

In the case of zero driving strength $\mu=0$, it follows $\Delta_l^\pm=0$ for all $l$, and $J_s(0)=\delta_{s,0}$.
The two dimensional lattice Eq.(\ref{eq:AM_newform}) turns into an infinite set of independent one-dimensional Anderson chains. 
For nonzero driving strength ($\mu\neq 0$) all the Floquet channels are connected.  
Since \cite{Watson22, Abramowitz72}
\begin{equation}
J_s(\Delta_l^\pm)\longmapsto \frac{(\Delta_l^\pm)^s}{2^s s! }\ ,\quad \text{for}\ \  s\longmapsto +\infty
\label{eq:Bessel_decay}
\end{equation}
only a finite number of channels will have a nonexponentially weak hopping strength, and that number depends on the value of the argument of the Bessel functions $\Delta_l^\pm$. If for every $l$ the argument is close to zero $|\Delta_l^\pm|\leq \delta \ll 1$, it follows that $J_0(\Delta_l^\pm)\approx 1$, and for $s\neq 0$ the Bessel functions $J_s(\Delta_l^\pm)$ decay as Eq.(\ref{eq:Bessel_decay}). As a consequence Eq.(\ref{eq:AM_newform}) becomes an infinite set of uncoupled equivalent Anderson chains, and the model resembles the unperturbed case. We call this the {\it single channel regime}.
This result is truly irrespective of the ratio of $\omega/W$ which controls the energy differences between connected sites. While this is evident for large frequencies $\omega \gg W$ 
(where the energy difference is of order $W$), it is also true in the opposite limit $\omega \ll W$ where the energy difference can be of order $\omega$ for a suitably large value of $s \sim W/\omega$, since the hoppings decay superexponentially fast with large $s$ (\ref{eq:Bessel_decay}).

In contrast, when $\Delta_l^\pm \geq 1$, a finite number $\mathcal{L}_l^\pm$ of the Bessel functions $J_s(\Delta_l^{\pm})$ for $s\neq 0$ are not negligible. Their typical values  can be approximated as \cite{Watson22, Abramowitz72}:
\begin{equation}
J_s(\Delta_l^\pm)\approx \sqrt{ \frac{2}{\pi \Delta_l^\pm} }\cos\bigg(\Delta_l^\pm-\frac{s \pi}{2} - \frac{\pi}{4} \bigg)\ ,\quad |s|\leq \frac{\mathcal{L}_l^\pm }{2}
\label{eq:Bessel_approximation_s}
\end{equation}
In Eq.(\ref{eq:sqrt_Rphases}), the square root of $\Delta_l^\pm$ is on the order of the disorder strength $W$. We define the parameter $\Delta$ as
\begin{equation}
\Delta_l^{\pm} \quad\longmapsto\quad \Delta= \frac{\mu W}{\omega} \ ,
\label{eq:threshold_1}
\end{equation}
It follows that, if $\Delta \geq 1$, the averaged number of connected channels $\mathcal{L}$ is given by 
\begin{equation}
\mathcal{L}_l^\pm= 2\Delta_l^\pm \quad\longmapsto\quad \mathcal{L}= 2\Delta = \frac{2\mu W}{\omega} \ ,
\label{eq:threshold_L}
\end{equation}
while in Eq.(\ref{eq:Bessel_approximation_s}), neglecting the cosine term that represents the Bessel function oscillation, the general value of the hopping can be approximated as
\begin{equation}
J_s \approx \sqrt{ \frac{2}{\pi \Delta} }= \sqrt{\frac{2\omega }{\pi \mu W} }\ ,\quad |s|\leq \frac{\mathcal{L}}{2}
\label{eq:Bessel_approximation_s2}
\end{equation}
We call this regime the {\it multi-channel regime}. Therefore, for any given disorder and driving strengths $W$ and $\mu$, this regime will be realized  in the limit of small frequencies $\omega$. This latter regime will be the focus of our investigation, since in the single channel regime the model acts similar to the undriven case.

\subsection{The multi-channel regime} \label{}

\subsubsection{Weak driving} \label{sec:1Dwd}

Let us first consider the case $\mu \ll 1$. For a given site $(l,\nu)$ there are $\mathcal{L}$ matrix elements connecting this site to a set of sites $(l+1,\nu')$. The onsite energies of these
connected sites vary in an interval $[\epsilon_{l+1}-\mu W, \epsilon_{l+1} + \mu W]$. For $\mu \ll 1$ this interval is narrow compared to $W$ which characterizes the spread of $\epsilon_l$.
Therefore the difference in the onsite energies between two connected sites is still of the order $W$.

In general, for ladders with a finite number $N$ of equivalent Anderson chains, Dorokhov, Mello, Pereyra and Kumar \cite{Dorokhov83, Mello88} estimate 
the localization length $\zeta$ of a $N$-leg ladder as a product of the localization length $\zeta_{\mathcal{A}}$ of each leg and the number of legs $N$:
\begin{equation}
\zeta \sim N\cdot \zeta_{\mathcal{A}}\;.
\label{eq:dorokhov}
\end{equation} 
The number of legs $N$ corresponds to $\mathcal{L}$. The single channel localization length can be estimated using the ratio $W_{\text{eff}}$ between the energy mismatch (disorder strength) $W$ and
the matrix element $\lambda J_s$ from (\ref{eq:Bessel_approximation_s2}):



\begin{equation}
W_{\text{eff}} = \frac{W^{\frac{3}{2}}}{\lambda } \sqrt{ \frac{\mu\pi}{2\omega}   }
\;.
\label{weff}
\end{equation}
We then obtain \cite{Kramer93}
\begin{equation}
\zeta_\mathcal{A}(\omega) \sim
 \left\{\begin{array}{rl}
\frac{100}{W_{\text{eff}}^2}  \quad\ \ &\text{if } \ \ W_{\text{eff}} \leq  10,\\
\frac{1}{|\ln(1/W_{\text{eff}})|}  \quad&\text{if }\ \  W_{\text{eff}} >10.
\end{array}  \right.
\label{eq:eff loc length}
\end{equation}
It follows that in the case of weak effective disorder $W_{\text{eff}}\leq10$, the localization length $\zeta$ of our driven model is given by
\begin{equation}
\zeta(\omega)
\sim \frac{400\lambda^2}{\pi W^2}\ \;,\; \omega  \geq \frac{\mu \pi W^3}{200 \lambda^2}\;.
\label{eq:eff loc length_D=1,We<10}
\end{equation}
The localization length $\zeta$ does not depend on the frequency $\omega$ and the driving strength $\mu$, since the increase of the number of connections $\mathcal{L}$ and the decrease of the localization length $\zeta_\mathcal{A}$ along each Anderson chain balance each other. Upon further decrease of the frequency $\omega$, this balancing effect is destroyed  
since $\Delta$ grows, the matrix element $\lambda J_s$ decays, and the effective disorder $W_{\text{eff}} > 10$. 
With (\ref{eq:eff loc length}) it follows
\begin{equation}
 \zeta(\omega)=\frac{2\mu W}{\omega}\cdot \Bigg| \ln\bigg(\frac{\lambda }{W^{\frac{3}{2}}} \sqrt{ \frac{2}{\pi\mu}\omega   }  \bigg)   \Bigg|^{-1} \;,\;
\omega  \leq \frac{\mu \pi W^3}{200 \lambda^2}\;.
\label{eq:eff loc length_D=1,We>10}
\end{equation}
The localization length then diverges for $\omega \rightarrow 0$. 

To summarize: in the weakly driven multichannel regime we expect a plateau in the dependence $\zeta(\omega)$  for $\omega  \geq \frac{\mu \pi W^3}{200 \lambda^2}$,
which is replaced by a divergence for $\omega  \leq \frac{\mu \pi W^3}{200 \lambda^2}$.




\subsubsection{Strong driving}

Let us consider the case $\mu \geq 1$, and constant phases $\phi_l=\text{const}$. For a given site $(l,\nu)$ there are $\mathcal{L}$ matrix elements connecting this site to a set of sites $(l+1,\nu')$. The onsite energies of these
connected sites vary in an interval $[\epsilon_{l+1}-\mu W, \epsilon_{l+1} + \mu W]$. For $\mu \geq 1$ this interval is larger than $W$ which characterizes the spread of $\epsilon_l$.
Therefore there will be typically one onsite energy amongst the connected set which is detuned by a mismatch of order $\omega$ from the one on site $(l,\nu)$. 
It follows that for $\mu\geq 1$ and $\omega \ll W$ we can trace a path in Eq.(\ref{eq:AM_newform}) where the onsite mismatch is of the order of the frequency $| \epsilon_l - \epsilon_{l\pm1} + s \omega| \sim \omega$ and the hopping scales with the square root of the frequency $J_{s}\sim\sqrt{\omega}$ for every $l$.  We call this the {\it optimal path}. The optimal path is a one dimensional random walk within the two dimensional network. 
Along that optimal path the localization length $\zeta_\mathcal{OP}$ follows from the effective disorder
$W_{\text{eff}}$ between the energy mismatch (disorder strength) $\omega$ and
the matrix element $\lambda J_s$ from (\ref{eq:Bessel_approximation_s2})
  \begin{equation}
\zeta_\mathcal{OP}(\omega)=\frac{100 J_{\hat{s}}^2}{\omega^2} = \frac{100}{\omega^2}.\Bigg(\sqrt{\frac{2\omega}{\pi\mu W}}\Bigg)^2=\frac{200}{\pi\mu W}\frac{1}{\omega}\ .
\label{eq:opt path loc length}
\end{equation}
For small frequencies this optimal path consists of strong links. Furthermore, paths neighboring the optimal one, are also strong links, as long as the energy detuning is not exceeding the matrix
element. Since the matrix element scales with $\sqrt{\omega}$, the number of strong links diverges as $N \sim 1/\sqrt{\omega}$. Using the Dorokhov estimates, we conclude that
the localization length on the network is scaling as
\begin{equation}
\zeta \sim \frac{200}{\pi\mu W}\frac{1}{\omega^{3/2}}\ .
\label{eq:loc length OP}
\end{equation}
Therefore the localization length diverges for vanishing frequency faster than in the weak driving case.

\subsubsection{Local suppression of strong driving}

The presence of random phases will suppress the 
optimal path through an increase of the minimal value of the driving strength:

\begin{equation}
\mu\sqrt{ 1 + \frac{ 2\epsilon_l \epsilon_{l\pm1} }{( \epsilon_l - \epsilon_{l\pm1})^2}   (1- \cos \theta_l^\pm   )} \geq 1
\label{eq:mu>1 general equivalent}
\end{equation}
where $\theta_l^\pm = \phi_l - \phi_{l\pm 1}$. In particular if $ \theta_l^\pm =\pi$, the square root term of Eq.(\ref{eq:mu>1 general equivalent}) is equal to zero if 
\begin{equation}
-2\epsilon_l \epsilon_{l\pm1} =  \epsilon_l^2 + \epsilon_{l\pm1}^2 \quad\Leftrightarrow\quad     \epsilon_l = -  \epsilon_{l\pm 1}
\label{eq:mu>1 eps1}
\end{equation}
In this case, Eq.(\ref{eq:mu>1 general equivalent}) does not hold for any finite value of $\mu$ and locally between site $l$ and site $l\pm 1$ the optimal path is not accessible.

Therefore, for uncorrelated phases $\phi_l$, the square root term of Eq.(\ref{eq:mu>1 general equivalent}) can be arbitrarily close to zero if $  \theta_l^\pm \approx\pi$ and $ \epsilon_l \approx -  \epsilon_{l\pm 1}$. As a consequence, $\mu$ has to diverge to infinity in order to satisfy Eq.(\ref{eq:mu>1 general equivalent}) at every step $l$ and so, for given finite values of the driving strength, the optimal path is interrupted. This will lead to a slower divergence of the localization length, which however is still expected to be faster than in the weak driving regime, since 
there will be finite volume parts in which the optimal path will survive.

\subsubsection{Local suppression of the multi-channel regime}


Since $\epsilon_l$ and $\phi_l$ are random phases, the square root term of Eq.(\ref{eq:sqrt_Rphases}) can be arbitrarily close to zero. Therefore, even deep in the multi-channel regime $\Delta\gg 1$, there might exist one or more lattice sites $l$ such that the Bessel function argument $\Delta_l^\pm \ll 1$. In that case, the hopping $J_s(\Delta_l^\pm)\approx\delta_{s,0}$ as in the single channel regime. As a result, the multi-channel regime between site $l$ and $l\pm 1$ becomes locally suppressed.


In the case of constant phases $\phi_l=\text{const}$, this local suppression appears if $| \epsilon_l -  \epsilon_{l\pm 1}|\ll\omega/\mu$.
The single channel then 
still shows an energy difference of the order of $\omega$ or less, similar to the optimal path.
For uncorrelated phases $\phi_l$, the probability of a local suppression of the multi-channel regime is reduced. Indeed, in order to violate the multi-channel condition 
$\Delta_l^{\pm} \gg 1$ we now need request either $|\epsilon_l - \epsilon_{l+1}| \ll \omega^2/\mu^2 $ and $|\theta_l^{\pm}| \ll 1$ or
$|\epsilon_l + \epsilon_{l+1}| \ll \omega^2/\mu^2$ and $ ||\theta_l^{\pm}|-\pi | \ll 1$. 

The previously obtained estimates on the localization length $\zeta$ in either weak and strong driving regimes are therefore upper bounds.

\section{Many colors} \label{sec:General case}

For the general case of Eq.(\ref{Eq:multicolor}), the Floquet expansion in the momentum space and the rotation of the eigenvalue problem in a basis of Bessel functions for each
frequency is a natural generalization of what we have previously described in Sec.\ref{sect:D=1} for one color $D=1$ (see appendix). 
Since the frequency components of $\Omega$ are chosen to be incommensurate Eq.(\ref{eq:incommens}), the general form of the Floquet expansion (\ref{eq:tr2}) runs over the vector index ${\bf k}=(k_1,\dots,k_D)\in\mathbb{Z}^D$ \cite{Kim88}, 
and yields a $D+1$ dimensional time independent eigenvalue problem:
\begin{equation}
\begin{split}
&E c_{l,{\bf v}} = (\epsilon_l + \Omega\cdot{\bf v})c_{l,{\bf v}}  \\
-\lambda   \sum_{{\bf s}} &
\Big[   {\rm e}^{i {\bf s}\cdot \Phi_l^-}    \mathcal{J}_{\bf s}^-     c_{l - 1,{\bf v} - {\bf s}}  +  {\rm e}^{i {\bf s}\cdot \Phi_l^+}   \mathcal{J}_{\bf s}^+       c_{l + 1,{\bf v} - {\bf s}} \Big] \;,
\end{split}
\label{eq:AM_newform_general case}
\end{equation}
where ${\bf s}, {\bf  v}\in \mathbb{Z}^D$ and $ \Phi_l^\pm = (\varphi_{l,i}^\pm)_{i=1}^D$. The coefficients are defined as 
\begin{equation}
\begin{split}
& \tan \varphi_{l,i}^{\pm}  = -\frac{\epsilon_{l\pm1} \sin \theta_{l,i}^{\pm}}{\epsilon_l - \epsilon_{l\pm1}\cos \theta_{l,i}^{\pm}} \;, \\ 
&\mathcal{J}_{\bf s}^\pm  =    \prod_{i=1}^D   J_{ s_i}\big(  \Delta_{l,i}^{\pm}  \big)\ ,\\
&\Delta_{l,i}^{\pm}  = \frac{\mu_i}{\omega_i}\sqrt{ \epsilon_l^2 + \epsilon_{l\pm1}^2 - 2\epsilon_l \epsilon_{l\pm1}\cos  \theta_{l,i}^{\pm}   }\;,
\end{split}
\label{eq:sqrt_Rphases_general case}
\end{equation}
which depend on the random phase difference $ \theta_{l,i}^{\pm}= \phi_l^i - \phi_{l\pm 1}^i$ for $i=1,\dots, D$. The resulting eigenvalue problem Eq.(\ref{eq:AM_newform_general case}) has $D$ frequency (color) directions (with zero hopping along them) and each site $c_{l,{\bf v}}$ is connected to the nearest neighbor ones $c_{l\pm1,{\bf v} - {\bf s}}$ through the complex matrix elements $ {\rm e}^{i {\bf s}\cdot \Phi_l^\pm}  \mathcal{J}_{\bf s}^\pm $. 
Each frequency color will add to the total number of channels. 

Let us assume that all frequency components satisfy the same condition for single or multi-channel (either strong or weak driving) regimes.
Then we conclude that the single channel regime will be applicable to multi-color driving as well, i.e. in the limit of large frequencies localization length corresponds to its value from the undriven case.
In the multi-channel regime, using the above argument of Dorokhov, the localization length $\zeta_D$ will be of the order of $\zeta_D \sim D \zeta$ with $\zeta$ being the corresponding single color localization length discussed in the previous section.
Therefore the localization length will stay finite for any finite number of colors. The divergence of $\zeta_D$ in the limit of $D \rightarrow \infty$
is in agreement with the result that a random noise driving leads to dephasing and complete delocalization \cite{Rayanov13}. 

\section{Numerical results} \label{sect:wave dyn}

We first analyze the single color driving and then discuss the two color case. We decide not to diagonalize the eigenvalue problem Eq.(\ref{eq:AMrot1}), since this will limit the required system size $N$ and the number of colors $D$. Instead, we compute the spreading of the wave packet over time for a single site excitation $\psi_{l}(t=0) = \delta_{l,N/2}$   
as an initial condition using a numerical symplectic integration scheme $\text{SBAB}_2$ \cite{Laskar01,symplectic}. To measure the evolution of the wave packet we calculate the second moment $m_2$ which, for localized modes, estimates the squared distance between the eigenmode tails. It is related to the localization length $\zeta$ of one mode as $m_2\sim \zeta^2$, and is defined as
\begin{equation}
m_2 (t) = \sum_l \left( l  -\sum_{l'}  l' \big|\psi_{l'} (t)\big|^2 \right)^2  \big|\psi_l (t)\big|^2\ .
\label{eq:m2_definition}
\end{equation}
Hereafter, unless indicated differently, in Eq.(\ref{Eq:multicolor}) we choose the hopping amplitude $\lambda=1$ and the disorder strength $W=4$. Furthermore, for the numerical computations we choose a system size $N=1024$ and we average $\log_{10}m_2(t)$  over $512$ disorder realizations, unless stated otherwise.

\subsection{One color }

\begin{figure}[h]
  \begin{center}
      \includegraphics[width=1.1\columnwidth]{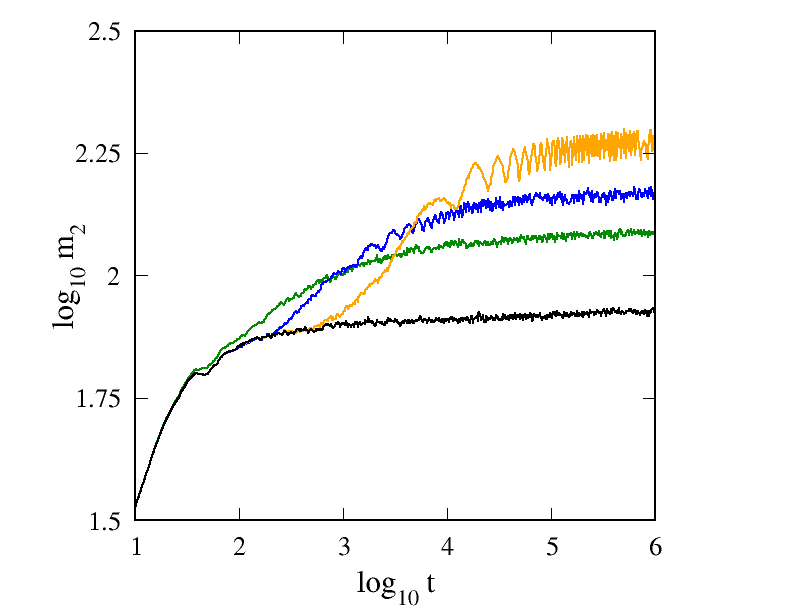}
    \caption{Time evolution of the second moment $m_2$ with driving strength $\mu=0.1$. Frequency values (from top to bottom at the right edge of the plot): 
$\omega=5\cdot 10^{-4}$ (orange),
$\omega=5\cdot 10^{-3}$ (blue),
$\omega=10^{-1}$ (green), 
undriven (black).
}
    \label{fig:M2_vs_t_1e6}
  \end{center}
\end{figure}

Let us analyze the frequency dependence of the time evolution. For the {\it weak driving} case we choose the driving strength $\mu=0.1$. The multi-channel regime ($\Delta\geq 1$) is then obtained for frequencies $\omega\leq 0.4$. The time evolution of the second moment $m_2$ is shown in Fig. \ref{fig:M2_vs_t_1e6}. 

We observe that the second moment first increases with time, but saturates at later times, indicating a halt of spreading, and a localization of the wave packet. Therefore we conclude that
there is a finite upper bound on the localization length $\zeta$ of the corresponding eigenvalue problem. The onset of spreading beyond the undriven reference curve (horizontal black one in 
 Fig.~\ref{fig:M2_vs_t_1e6}) scales inversely with the driving frequency as expected.

\begin{figure}[h]
  \begin{center}
      \includegraphics[width=1.1\columnwidth]{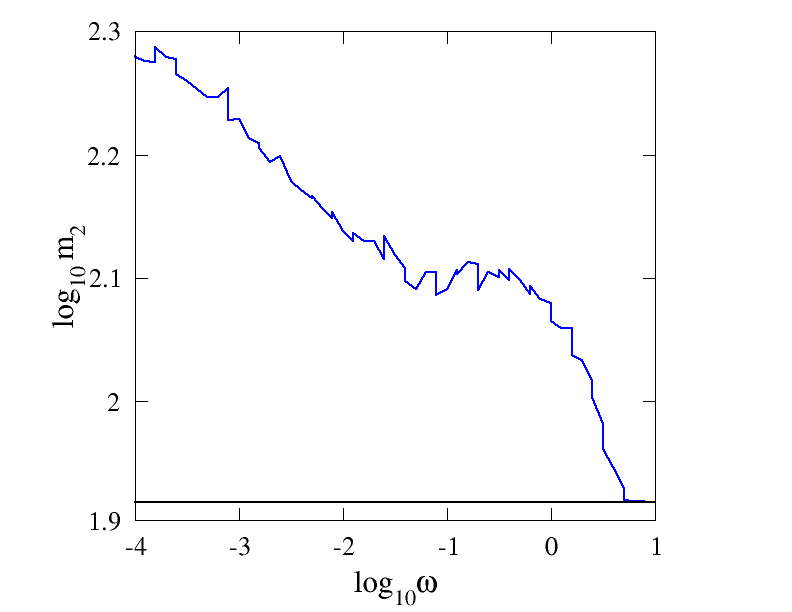}
  \caption{Saturated second moment $m_2$ at $t=10^6$ as a function of frequency $\omega$ of a one-color perturbation with $\mu=0.1$. The black horizontal line indicates the value for the undriven case. 
  } 
    \label{fig:mu2_vs_omega_tmax1e6_mu1tenth}
  \end{center}
\end{figure}

We use the saturated value of the second moment at time $t=10^6$ as a measure of the squared localization length and plot it as a function of frequency $\omega$ in
Fig.\ref{fig:mu2_vs_omega_tmax1e6_mu1tenth} .
Increasing $\omega$ in the single channel regime $\omega > 0.4$ leads to a quick decay of the saturated second moment to reach the reference value of the undriven case already at $\omega \approx 4$. In the multichannel regime we observe two features: a plateau at an intermediate frequency interval, and a subsequent increase of the saturated second moment by further lowering the frequency. This is in agreement with our analytics in Sec. \ref{sec:1Dwd}.

\begin{figure}[h!]
  \begin{center}
          \includegraphics[width=1.1\columnwidth]{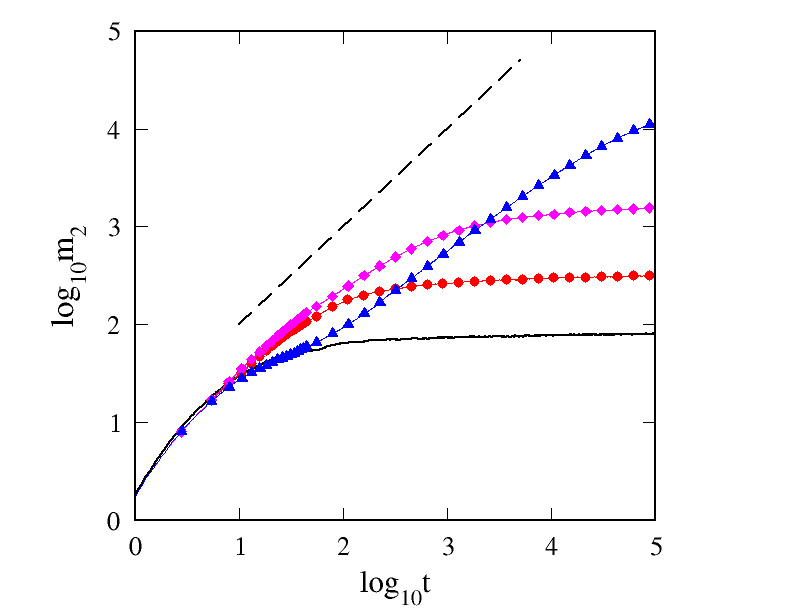}
          \caption{Time evolution of the second moment $m_2$ with driving strength $\mu=1$. The dashed line indicates normal diffusion $m_2 \sim t$.  Frequency values: $\omega=5$ (black), $\omega=2$ (red/circles), $\omega=0.5$ (magenta-diamonds), $\omega=0.02$ (blue-triangles).}
    \label{fig:m2_vs_time_mu_one} 
  \end{center}
\end{figure}
\begin{figure}[h!]
  \begin{center}
          \includegraphics[width=1.1\columnwidth]{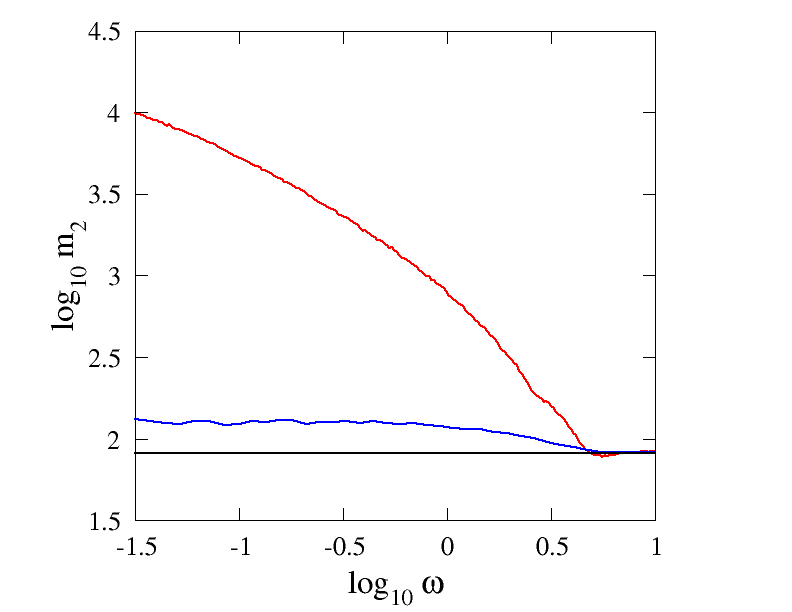}
          \caption{ Saturated second moment $m_2$ at $t=10^5$ as a function of frequency $\omega$ of a one-color perturbation with $\mu=0.1$ (blue, bottom) and $\mu=1$ (red, top). The black horizontal line indicates the value of the undriven case.}
    \label{fig:m2_vs_omega_mu_one}
  \end{center}
\end{figure}

For the {\it strong driving} regime, we consider a driving strength $\mu=1$. In Fig.~\ref{fig:m2_vs_time_mu_one} we plot the time evolution of the second moment $m_2$ for different frequencies in the multi-channel regime $\omega\leq 5$.  
The second moment $m_2$ increases as the frequency $\omega$ decreases, in agreement with Eq.(\ref{eq:loc length OP}). Moreover, we observe the appearance of transient regions of normal diffusion that extend their length as the frequency $\omega$ decreases. Again $m_2(t)$ saturates at larger time, indicating a halt of spreading, and a localization of the wave packet.
The frequency dependence of the saturated values of the second moment $m_2$ is shown in Fig.~\ref{fig:m2_vs_omega_mu_one}, where we plot the second moment of the wave packet at time $t=10^5$ as function of the frequency $\omega$.  For comparison we also replot the weak driving curve from Fig.~\ref{fig:mu2_vs_omega_tmax1e6_mu1tenth}.  
Similar to the weak driving (blue curve), the second moment for the strong driving (red curve) tends to the undriven case (black horizontal line) for large frequencies and diverges for small ones. 
 In agreement with Eq.(\ref{eq:loc length OP}), the plateau (which was observed for weak driving) is suppressed, although a reminding shoulder exists at $\omega\approx 2.5$.
The strong driving yields much larger values of the saturated second moment as compared to the weak driving case, in accord with our predictions.
The frequency dependence is weaker than
the predicted law in Eq.(\ref{eq:loc length OP}), most likely due to the discussed local suppression of strong driving and multi-channel regimes.

\begin{figure}[h]
  \begin{center}
          \includegraphics[width=1.1\columnwidth]{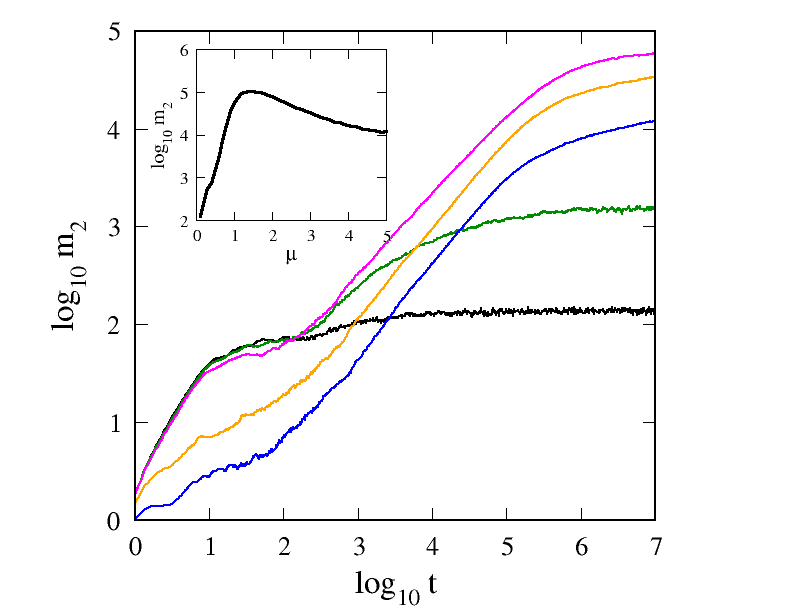}
          \caption{Time evolution of the second moment $m_2$ with frequency $\omega=5\cdot 10^{-3}$. Driving strength values (from top to bottom at the right edge of the plot): 
$\mu=1.0$ (magenta),
$\mu=3.0$ (orange),
$\mu=5.0$ (blue),
$\mu=0.5$ (green), 
$\mu=0.1$ (black). The inset plot shows the dependency of the saturated second moment $m_2$ at $t=10^7$ on the drive strength. System size $N=2048$, and log$_{10}$m$_2$ is averaged over 32 disorder realizations.
}
    \label{fig:m2_vs_mu}
  \end{center}
\end{figure}

In Fig.~\ref{fig:m2_vs_mu} we plot the time evolution of the second moment $m_2$ for different values of the driving strength $\mu$ in the strong driving regime. The inset of Fig.~\ref{fig:m2_vs_mu} shows the dependence of the saturated second moment $m_2$ at $t=10^7$ on $\mu$. We observe that the saturated moment increases up to $\mu\approx 1$ and starts to decrease for larger values of $\mu$.
This subsequent decrease is qualitatively in agreement with the prediction $\zeta\sim\frac{1}{\mu}$ following from Eq.(\ref{eq:AM_newform}).

\subsection{Two colors }

\begin{figure}[h]
  \begin{center}
          \includegraphics[width=1.1\columnwidth]{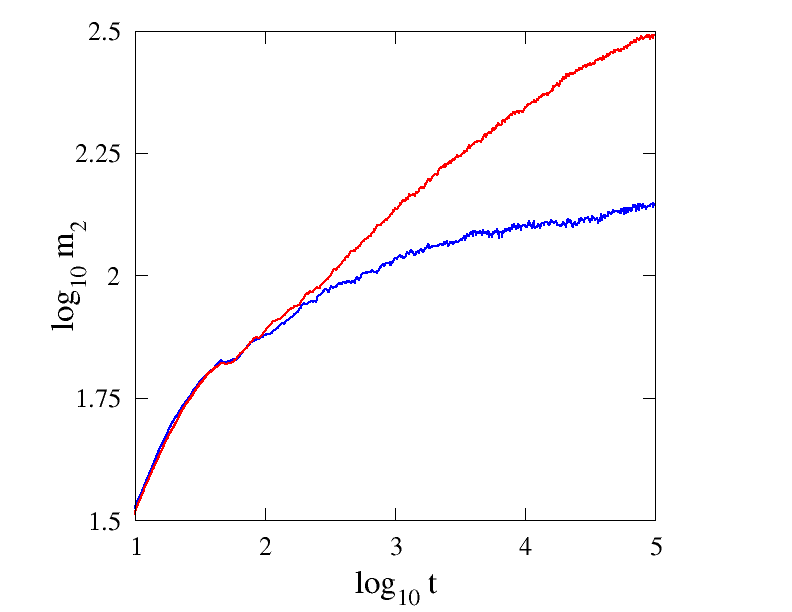}
          \caption{Time evolution of the second moment $m_2$ for one-color (blue, bottom) and two-color (red, top) cases with driving strength $\mu=\mu_1=0.1$ and frequency 
$\omega=\omega_1=3\cdot 10^{-2}$. We recall $\omega_2=\sqrt{2}{\omega_1}$ for the two color case. }
    \label{fig:2color_compare}
  \end{center}
\end{figure}

We assume the driving strengths to be equal $\mu_1=\mu_2$ and we fix the frequency relation $\omega_2=\sqrt{2}\omega_1$. We first consider the {\it weak driving} case. In Fig. \ref{fig:2color_compare}, we compare one and two color cases for same driving strength $\mu =0.1 =\mu_1$ and frequency $\omega=3\cdot 10^{-2}=\omega_1$. 

The presence of a second incommensurate driving term enhances the spreading of the wave packet. However, the integration time $t=10^5$ is not enough to see the saturation of the second moment. 

\begin{figure}[h!]
  \begin{center}
          \includegraphics[width=1.1\columnwidth]{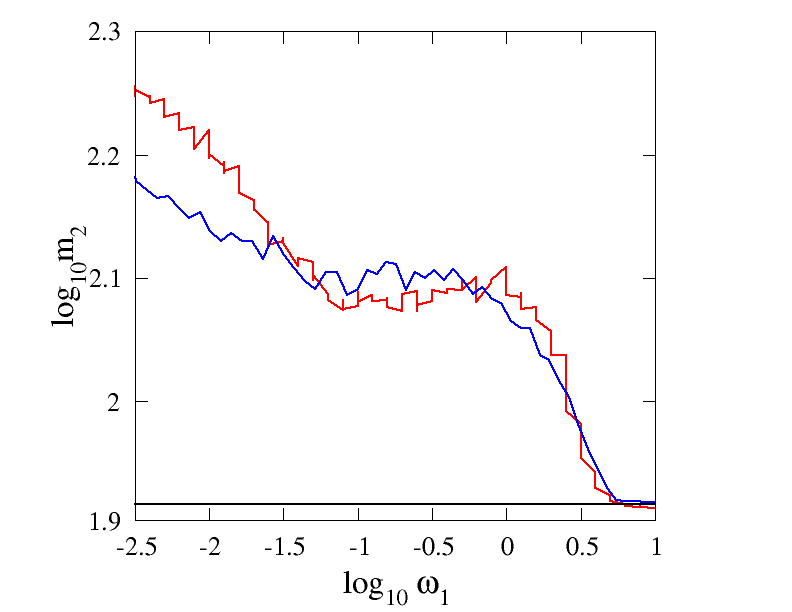}
          \caption{Saturated second moment $m_2$ at $t=10^5$ as a function of driving frequency $\omega_1$ with driving strengths $\mu_1=\mu_2=0.05$ (red curve, top at left corner). 
The blue curve corrsponds to the single color one from Fig.\ref{fig:mu2_vs_omega_tmax1e6_mu1tenth}.
We recall $\omega_2=\sqrt{2}{\omega_1}$. The black horizontal line indicates the value of the undriven case.
          }
    \label{fig:m2_vs_omega_2color}
  \end{center}
\end{figure}

To obtain saturation at time $t=10^5$, we reduce the driving strength to $\mu_1=0.05=\mu_2$. In Fig.\ref{fig:m2_vs_omega_2color} we plot the saturated value of the second moment $m_2$ at time $t=10^5$, as a function of the frequency $\omega_1$ and compare it to the weak driving single color result ($\mu=0.1$) from Fig.\ref{fig:mu2_vs_omega_tmax1e6_mu1tenth}.
The values approach the undriven case (black horizontal line) for approximately $\omega_1\geq 4$, in good agreement with the single channel regime 
$\omega_1,\omega_2\geq 2$. 
Similar to the one color case, the saturated second moment exhibits a plateau in the multi-channel regime. 
Notably the height of the plateau is practically equal to the single color one reported in Fig.\ref{fig:mu2_vs_omega_tmax1e6_mu1tenth}, in full accord with our estimate of the localization
length in Eq.(\ref{eq:eff loc length_D=1,We<10}), which is independent of the driving strength $\mu$ and more general independent of the number of participating channels. Therefore
the presence of a second frequency which increases the number of channels, should not change the plateau value, as observed.

\begin{figure}[h!]
  \begin{center}
      \includegraphics[width=1.1\columnwidth]{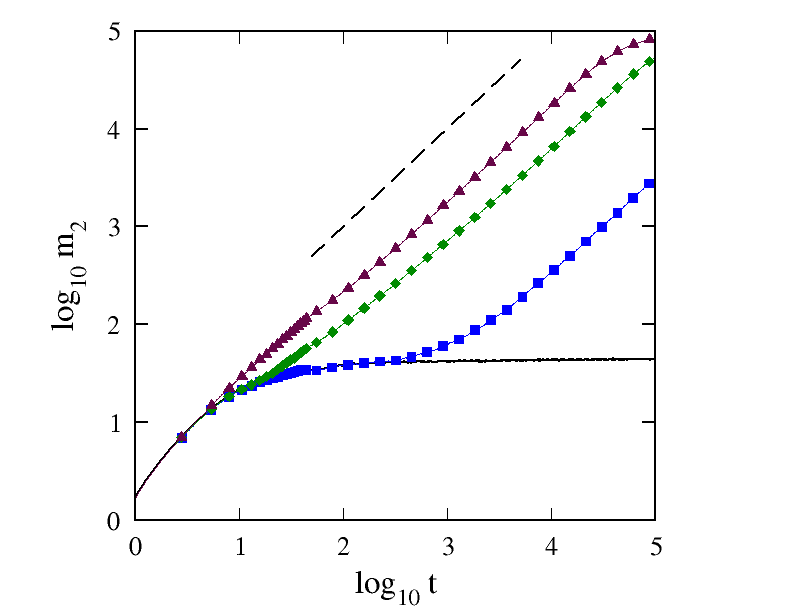}
    \caption{
Time evolution of the second moment $m_2$ with driving strength $\mu_1=\mu_2=1$. We recall $\omega_2=\sqrt{2}{\omega_1}$. The dashed line indicates normal diffusion $m_2 \sim t$.  Frequency values: $\omega_1=5\cdot 10^{-1}$ (maroon/triangles), $\omega_1=5\cdot 10^{-2}$ (green/diamonds), $\omega_1=5\cdot 10^{-4}$ (blue/squares), undriven (black).}
    \label{fig:2color_diff_mu}
  \end{center}
\end{figure}

For the {\it strong driving} case, we consider $\mu_1=\mu_2=1$. In Fig. \ref{fig:2color_diff_mu} we show the time evolution of the second moment $m_2$ for frequencies $\omega_{1,2}$ chosen in the multi-channel regime. 
We observe long lasting transient regions of diffusive spreading, with a subsequent halt and localization. 
We also observe a significant increase in the localization length as compared to the single color case, in agreement with our predictions.
\section{Summary and Conclusions}

In this work we have studied the spreading of a wave packet for a one-dimensional disordered chain in the presence of a multi-frequency quasi-periodic drive. For each term (color) of the driving, the Floquet representation is used to arrive at a time-independent eigenvalue problem on a two-dimensional
lattice, with one direction corresponding to the original spatial extension, and the second one to the Floquet (driving) one. We transform into a Wannier-Stark basis which is diagonal along
the Floquet direction, and analyze the resulting eigenvalue problem. For large driving frequencies the equations reduce to uncoupled single channel ones, which are essentially equivalent to
the undriven case. For small driving frequencies we obtain a multi-channel regime with a substantial increase of the localization length, and its divergence in the limit of vanishing frequency.
This multi-channel regime divides into two further regimes of weak and strong driving amplitudes, which yield different scaling laws.

In the many colors case, we have shown that each incommensurate color is independent from the others and can be treated separately.
The localization length of the model will then be proportional to the single color localization length, and scales with the number of colors set in the multi-channel regime. Therefore, it will remain finite for any finite number of colors. 

Numerically we have observed that in the strong driving, the model exhibit transient regions of diffusive dynamics before localization occurs. The localization volume
increases as the number of colors increases, when satisfying the multi-channel regime condition.
It follows that for $D \rightarrow \infty$, the localization length will diverge to infinity and the region of diffusive dynamics will extend to infinity until a complete delocalization 
is observed. The divergence of the number of colors $D$ corresponds to the loss of quasiperiodicity of the driving term, and consequently to an effective random driving which leads to 
a loss of Anderson localization in that limit.

Mathematical studies of the single color $D=1$ case \cite{Soffer03} and multicolor case $D > 1$ \cite{Bourgain04} predict stability of Anderson localization in the regime of strong disorder.
These results are in line with our findings, since the limit of strong disorder is corresponding to the single channel regime, which is essentially independent on the number of colors, 
with a localization length close to the one of the undriven case.

\appendix*
\section{Floquet analysis and coordinate transformation} \label{sect:appendix}

We focus on the one color case $D=1$ using references \cite{Watson22, Abramowitz72}. The $D$ color case is a generalization of these calculations. The one-dimensional time-dependent model Eq.(\ref{Eq:onecolor}) 
is mapped to a time independent two-dimensional eigenvalue problem Eq.(\ref{eq:AMrot1}) via the Floquet expansion Eq.(\ref{eq:tr1})
\begin{equation}
\psi_l (t)=  \sum_k A_{l,k} e^{-i[(E- k\omega ) t - k \phi_l]}\;.
\label{eq:tr1_app}
\end{equation}
With
\begin{equation}
\begin{split}
&\quad\mu \epsilon_l \cos(\omega t + \phi_l) \psi_l = \\
&= \frac{\mu\epsilon_l}{2} \sum_k \big[ A_{l,k-1} + A_{l,k+1}\big]e^{-i[(E-k\omega) t - k\phi_l]} 
\end{split}
\label{eq:driving exp appendix}
\end{equation}
and
\begin{equation}
\begin{split}
&\quad \lambda(\psi_{l+1}+\psi_{l-1}) = \\
&=\lambda \sum_k \Big[A_{l-1,k} e^{-i[(E- k\omega ) t - k \phi_{l-1}]} \\
&\qquad \quad+  A_{l-1,k} e^{-i[(E- k\omega ) t - k \phi_{l-1}]}\Big] \\
& - \lambda\sum_k   \Big[A_{l-1,k} e^{-i k(\phi_l -\phi_{l-1})} e^{-i[(E-k\omega) t - k \phi_{l}]}  \\
&\qquad\quad+ A_{l+1,k} e^{-i k(\phi_l - \phi_{l+1})}  e^{-i[(E-k\omega) t - k \phi_{l}]} \Big] \
\end{split}
\label{eq:hopping exp appendix}
\end{equation}
we define the hopping coefficients $\xi_{l,k}^{\pm}$ as 
\begin{equation}
\xi_{l,k}^{\pm} = e^{-i k(\phi_l -\phi_{l\pm 1})}
\label{eq:complex phase coefficients_app}
\end{equation}
to arrive at
\begin{equation}
\begin{split}
&\quad \sum_k  (E-k\omega ) A_{l,k}e^{-i[(E-k\omega ) t - k\phi_l]}= \\
&=  \sum_k \bigg[\epsilon_l A_{l,k} + \frac{\mu\epsilon_l}{2} \big( A_{l,k-1} + A_{l,k+1}\big)\\
& \qquad-\lambda \big(\xi_{l,k}^{-}  A_{l-1,k}   +  \xi_{l,k}^{+}  A_{l+1,k}  \big) \bigg] e^{-i[(E-k\omega) t - k\phi_l]}  \\
\end{split}
\end{equation}
which finally yields the two dimensional eigenvalue problem of Eq.(\ref{eq:AMrot1})
\begin{equation}
\begin{split}
E A_{l,k}& =( \epsilon_l + k \omega) A_{l,k} -\lambda (\xi_{l,k}^{-}  A_{l-1,k}   +  \xi_{l,k}^{+}  A_{l+1,k})\\
& + \frac{\mu\epsilon_l}{2}\big(  A_{l,k-1} +  A_{l,k+1}  \big)\ .
\end{split}
\label{eq:AMrot1_app}
\end{equation} 
This is then transformed using Eq.(\ref{eq:tr2}):
\begin{equation}
A_{l,k} = \sum_{\nu} c_{l,\nu}  B_{l,k}^{(\nu)}\ ,\qquad  B_{l,k}^{(\nu)} =  J_{k-\nu}\bigg(\frac{\mu\epsilon_l}{\omega} \bigg)\ .
\label{eq:tr2_app}
\end{equation}
The basis $\mathcal{B} = \{  B_{l,k}^{(\nu)}   \}_\nu$ diagonalizes the eigenvalue problem in the Fourier direction $k$ with eigenvalues $\epsilon_{\nu} = \omega\nu$. It follows that
\begin{equation}
\begin{split}
&\quad  k \omega A_{l,k} +  \frac{\mu\epsilon_l}{2}\big(  A_{l,k-1} +  A_{l,k+1}  \big)= \\
&  = \sum_{\nu} c_{l,\nu} \bigg[ k\omega   B_{l,k}^{(\nu)}+  \frac{\mu\epsilon_l}{2}\Big(  B_{l,k-1}^{(\nu)}  + B_{l,k+1}^{(\nu)}  \Big) \bigg] \\
& =  \sum_{\nu} \omega \nu\  c_{l,\nu}  B_{l,k}^{(\nu)} \;.
\end{split}
\label{eq:bessel diagAMrot1_app}
\end{equation} 
The eigenvalue problem Eq.(\ref{eq:AMrot1_app}) in the new basis $\mathcal{B}$ reads
\begin{equation}
\begin{split}
E \sum_{\nu}& c_{l,\nu}  B_{l,k}^{(\nu)} =\sum_{\nu}\Big[ (\epsilon_l + \omega\nu) c_{l,\nu}  B_{l,k}^{(\nu)}  \\
&-\lambda\big(\xi_{l,k}^{-} c_{l-1,\nu} B_{l-1,k}^{(\nu)} + \xi_{l,k}^{+} c_{l+1,\nu} B_{l+1,k}^{(\nu)}   \big) \Big]\ .
\end{split}
\label{eq:AMrot2A_app}
\end{equation}
We multiply both sides of Eq.(\ref{eq:AMrot2A_app}) with a second Bessel function $B_{l,k}^{(\nu_2)}$ of index $\nu_2$ and then sum over $k$.  Using the Bessel functions orthonormality relation \cite{Watson22, Abramowitz72} 
\begin{equation}
\sum_k B_{l,k}^{(\nu)}B_{l,k}^{(\nu_2)} = \sum_k J_{k-\nu}\bigg(\frac{\mu\epsilon_l}{\omega} \bigg) J_{k-\nu_2}\bigg(\frac{\mu\epsilon_l}{\omega} \bigg) = \delta_{\nu,\nu_2}   
\label{eq:Besselortho_app}
\end{equation}
in Eq.(\ref{eq:AMrot2A_app}) we obtain 
\begin{equation}
\begin{split}
&\qquad\quad E \sum_{k}\sum_{\nu} c_{l,\nu}  B_{l,k}^{(\nu)} B_{l,k}^{(\nu_2)}  =E c_{l,\nu}\ ,\\ 
&\sum_{k}\sum_{\nu}  (\epsilon_l + \omega\nu)c_{l,\nu}  B_{l,k}^{(\nu)} B_{l,k}^{(\nu_2)} = (\epsilon_l + \omega\nu)c_{l,\nu}\ .
\end{split}
\label{eq:AMrot2B_app}
\end{equation}
The matrix elements (hopping) along the real direction $l$ become
\begin{equation}
\begin{split}
&\quad\sum_{k}\sum_{\nu} \xi_{l,k}^{\pm}  c_{l\pm 1,\nu} B_{l \pm 1,k}^{(\nu)}B_{l,k}^{(\nu_2)} = \\ &=\sum_{\nu}  c_{l\pm 1,\nu}\bigg( \sum_{k}   \xi_{l,k}^{\pm} B_{l\pm 1,k}^{(\nu)}B_{l,k}^{(\nu_2)}\bigg)   \;.
\end{split}
\label{eq:AMrot2B3_app}
\end{equation}
Using Graf's generalization of Neumann's addition theorem \cite{Watson22, Abramowitz72} 
\begin{equation}
\begin{split}
\sum_k &  e^{ik\theta}J_{\nu +  k}(x) J_{k}(y) = \\
&=\bigg( \frac{x - y e^{-i\theta}}{ x - y e^{i\theta} }  \bigg)^{\frac{\nu}{2}} J_\nu \Big(\sqrt{x^2 + y^2 - 2xy\cos\theta}\Big) \;.
\end{split}
\label{eq:Bessel identity_app}
\end{equation}
and defining the random phase difference 
\begin{equation}
\theta_l^\pm = \phi_l -\phi_{l\pm 1}\ ,
\label{eq:phase difference_app}
\end{equation}
Eq.(\ref{eq:AMrot2B3_app}) is modified as
\begin{equation}
\begin{split}
&\sum_{k}\sum_{\nu} \xi_{l,k}^{\pm}  c_{l\pm 1,\nu} B_{l \pm 1,k}^{(\nu)}B_{l,k}^{(\nu_2)} = 
\sum_{s} \big( \sigma_l^{\pm} \big)^{  s}  J_{ s}\big(  \Delta_l^{\pm}  \big)  c_{l\pm 1,\nu-s}
\end{split}
\label{eq:AMrot2Bdone}
\end{equation}
where 
\begin{equation}
\begin{split}
\sigma_l^\pm &=\Bigg( \frac{\epsilon_l - e^{-i\theta_l^\pm}\epsilon_{l\pm 1} }{  \epsilon_l - e^{i\theta_l^\pm}\epsilon_{l\pm 1}}\Bigg)^{\frac{1}{2}} \;, \\
\Delta_l^\pm & = \frac{\mu}{\omega}\sqrt{ \epsilon_l^2 + \epsilon_{l\pm1}^2 - 2\epsilon_l \epsilon_{l\pm1}\cos \theta_l^\pm   } \;.
\end{split}
\label{eq:sqrt_Rphases_app}
\end{equation}
Since the complex coefficient $\sigma_l^\pm$ has absolute value equal to unity, we rewrite it as
\begin{equation}
\begin{split}
& \sigma_l^\pm = {\rm e}^{i\varphi_l^\pm}\ ,\qquad \tan \varphi_l^{\pm}  = -\frac{\epsilon_{l\pm1} \sin \theta_l^{\pm}}{\epsilon_l - \epsilon_{l\pm1}\cos \theta_l^{\pm}} \;.
\end{split}
\label{eq:sqrt_Rphases2_app}
\end{equation}
The final eigenvalue problem becomes (Eq.(\ref{eq:AM_newform}))
\begin{equation}
\begin{split}
&E c_{l,\nu} = (\epsilon_l + \omega\nu)c_{l,\nu} \\
-\lambda   \sum_{s} 
\Big[ & {\rm e}^{is\varphi_l^-}  J_{ s}\big(  \Delta_l^{-}  \big)  c_{l- 1,\nu-s} +  {\rm e}^{is\varphi_l^+}   J_{ s}\big(  \Delta_l^{+}  \big)  c_{l+ 1,\nu-s} \Big] \;.
\end{split}
\label{eq:AM_newform_app}
\end{equation}

\newpage



\begin{thebibliography}{99}
\bibitem{Anderson58} P.W. Anderson, Phys. Rev. {\bf 109}, 1492 (1958).
\bibitem{Abrahams79} E. Abrahams, P. W. Anderson, D. C. Licciardello, and T. V. Ramakrishnan, Phys. Rev. Lett. {\bf 42}, 673 (1979).
\bibitem{Bulka85} B.R. Bulka, B. Kramer and A. MacKinnon, Zeitschrift f{\" u}r Physik B Condensed Matter, {\bf 60}, 1, 13-17 (1985).
\bibitem{Lahini08} Y. Lahini, A. Avidan, F. Pozzi, M. Sorel, R. Morandotti, D.N. Christodoulides and Y. Silberberg, Phys. Rev. Lett. {\bf 100}, 013906 (2008)
\bibitem{Billy08} J. Billy, V. Josse, Z. Zuo, A. Bernard, B. Hambrecht, P. Lugan, D. Cl{\' e}ment, L. Sanchez-Palencia, P. Bouyer and A. Aspect, Nature {\bf 453}, 891-894 (2008). 
\bibitem{Roati08} G. Roati, C. D'Errico, L. Fallani, M. Fattori, C. Fort, M. Zaccanti, G. Modugno, M. Modugno and M. Inguscio, Nature {\bf 453}, 895-898 (2008).
\bibitem{Rayanov13} K. Rayanov, G. Radons and S. Flach, Phys. Rev. E {\bf 88} 012901 (2013).
\bibitem{Yamada93} H. Yamada, K. S. Ikeda, and M. Goda, Phys. Lett. A {\bf 182}, 77 (1993).
\bibitem{Yamada98} H. Yamada, and K. S. Ikeda, Phys. Lett. A {\bf 248}, 179 (1998).
\bibitem{Yamada99} H. Yamada, and K. S. Ikeda, Phys. Rev. E {\bf 59}, 5214 (1999).


\bibitem{Martinez06} D.F. Martinez, and R.A. Molina, Phys. Rev. B {\bf 73}, 073104 (2006).
\bibitem{Kitagawa12} T. Kitagawa,T. Oka, and E. Demler, Ann. Phys. {\bf 327}, 1868 (2012). 
\bibitem{Kramer93} B. Kramer and A. MacKinnon, Rep. Prog. Phys. {\bf 56}, 1469 (1993).
\bibitem{Floquet83} G. Floquet, Annales scientifiques De l'ENS, {\bf 12}, second edition, 47 (1883).
\bibitem{Shirley65} J. H. Shirley, Phys. Rev. {\bf 138}, B979 (1965).
\bibitem{Fukuyama73} H. Fukuyama, R. A. Bari, and H. C. Fogedby, Phys. Rev. B {\bf 8}, 5579 (1973).
\bibitem{Krimer09} D. O. Krimer, R. Khomeriki and S. Flach, Phys. Rev. E {\bf 80}, 036201 (2009).
\bibitem{Watson22} G.N. Watson, A Treatise on the Theory of Bessel function, Cambridge University press (1922).
\bibitem{Abramowitz72} M. Abramowitz and I. A. Stegun, {\it Handbook of Mathematical
Functions}, Dover Publications Inc., New York (1972).


\bibitem{Dorokhov83} O. N. Dorokhov, Solid State Commun. {\bf 46}, 605 (1983).
\bibitem{Mello88} P. A. Mello, P. Pereyra and N. Kummar, Ann. Phys. {\bf 181}, 290 (1988).
\bibitem{Kim88} S. Kim, S. Ostlund and G. Yu, Physica D {\bf 31} 117-126 (1988).
\bibitem{Laskar01} J. Laskar, and P. Robutel, Celest. Mech. Dyn. Astron. {\bf 80}, 39 (2001).
\bibitem{symplectic} Ch. Skokos, E. Gerlach, J.D. Bodyfelt, G. Papamikos and S. Eggl,  Phys. Lett. A {\bf 378} 1809 (2014);
E. Gerlach, S. Eggl, Ch. Skokos, J.D. Bodyfelt and G. Papamikos, Proc. of 10th HSTAM Intnl. Congress on Mechanics (2013);
E. Gerlach, J. Meichsner and Ch. Skokos,  arXiv:1512.07778.
\bibitem{Soffer03} A. Soffer,and W. Wang, Commun. Part. Diff. Eq. {\bf 28}, 333 (2003).
\bibitem{Bourgain04} J. Bourgain,and W. Wang, Commun. Math. Phys. {\bf 248}, 429 (2004).





%


\end{thebibliography}
\end{document}